\documentclass{aip-cp}

\usepackage[numbers]{natbib}
\usepackage{amssymb}
\usepackage{amsmath-aip}
\begin{document}

\title{Conservation laws and symmetries\\ of a generalized Kawahara equation}
\author[aff1]{Maria Luz Gandarias\corref{cor1}}
\author[aff1]{Maria Rosa }
\eaddress{maria.rosa@uca.es}
\author[aff1]{Elena Recio }
\eaddress{elena.recio@uca.es}
\author[aff2]{Stephen Anco}
\eaddress{sanco@brocku.ca}

\affil[aff1]{Departamento de Matematicas,  Universidad de Cadiz, Spain}
\affil[aff2]{Department of Mathematics and Statistics, Brock University, Canada}
\corresp[cor1]{Corresponding author: marialuz.gandarias@uca.es}

\maketitle

\begin{abstract}
The generalized Kawahara equation
$u_t=a(t) u_{xxxxx} +b(t)u_{xxx} +c(t)f(u) u_x$ 
appears in many physical applications.
A complete classification of low-order conservation laws and point symmetries
is obtained for this equation, 
which includes as a special case the usual Kawahara equation 
$u_t = \alpha u u_x+\beta u^2u_x  +\gamma u_{xxx}+\mu u_{xxxxx}$. 
A general connection between conservation laws and symmetries
for the generalized Kawahara equation is derived through 
the Hamiltonian structure of this equation 
and its relationship to Noether's theorem using a potential formulation. 
\end{abstract}

\section{Introduction}

Dispersive wave equations arise in many areas of applied mathematics and physics. 
One very important equation is the Korteweg-de Vries (KdV) equation
$u_t = \alpha u u_x+ \gamma u_{xxx}$
which models shallow water waves and is a bi-Hamiltonian integrable system. 
The dispersion is caused by the $u_{xxx}$ term, 
while the other term $uu_x$ describes nonlinear advection. 
A related equation is the modified KdV equation 
$u_t = \beta u^2 u_x+ \gamma u_{xxx}$ 
which has the same dispersion but a stronger nonlinearity. 
It is also a bi-Hamiltonian integrable system and models 
acoustic waves in anharmonic lattices \cite{Zab}
and Alfven waves in collision-free plasmas \cite{KakOno}. 

An interesting nonlinear wave equation that exhibits weaker dispersion than 
occurs for the KdV and modified KdV equations is the Kawahara equation 
\begin{equation}\label{Keq}
u_t = \alpha u u_x+\beta u^2u_x  +\gamma u_{xxx}+\mu u_{xxxxx} . 
\end{equation}
This a fifth-order dispersive nonlinear wave equation.  
It has a Hamiltonian structure, but in contrast to the KdV equation,
it is not an integrable system. 

The Kawahara equation has several important physical applications. 
It models plasma waves \cite{Kak,KakOno}
and capillary-gravity water waves \cite{Has}. 
Kawahara \cite{Kaw} studied this type of equation numerically 
and observed that it possesses both oscillatory and monotone solitary wave solutions.
Other basic aspects of the Kawahara equation are its symmetry structure 
and its set of conservation laws. 
Symmetries and conservation laws of particular cases of the Kawahara equation 
have been studied previously \cite{LiuLiLiu,FreSam}, 
but a complete classification of remains open. 

In the present paper, 
we consider a generalization of the Kawahara equation given by 
\begin{equation}\label{gKeq}
u_t = a(t) u_{xxxxx} + b(t)u_{xxx} + c(t)f(u) u_x
\end{equation}
where $f(u)$ is an arbitrary non-constant function, 
and $a(t),b(t),c(t)$ are arbitrary non-zero functions (which are allowed to be constant).
This generalization is able to model nonlinear weakly dispersive phenomena 
that involve a time-dependent coefficient. 
The Kawahara equation is a special case given by 
$a=\mu$, $b=\gamma$, $c=1$, $f=\alpha u + \beta u^2$. 

Our main results will be to provide a complete classification of all point symmetries 
and all low-order conservation laws admitted by 
the generalized Kawahara equation \eqref{gKeq}
in the case when $f$ is a nonlinear function. 
We will also discuss the Hamiltonian structure of this equation,
which provides a connection between the conservation laws and the symmetries. 

Note that, under a point transformation on $t$, 
all of the coefficients $a(t),b(t),c(t)$ in equation \eqref{gKeq}
can be multiplied by an arbitrary function of $t$. 
Consequently, hereafter we will assume 
\begin{equation}\label{classify}
a(t)=1,
\quad
b(t)\neq 0,
\quad
c(t)\neq 0,
\quad
f'(u)\neq 0 . 
\end{equation}

\section{Symmetries}

Symmetries are a basic structure as they can be used to find invariant solutions 
and yield transformations that map the set of solutions $u(t,x)$ into itself. 
A general discussion of symmetries and their applications to differential equations
can be found in Refs.\cite{Olv,BluCheAnc}. 

An infinitesimal point symmetry for the generalized Kawahara equation \eqref{gKeq} 
is a generator 
\begin{equation}
\mathrm{X} =\xi(x,t,u)\partial_x+\tau(x,t,u)\partial_t+\eta(x,t,u)\partial_u
\end{equation} 
whose prolongation leaves invariant the equation \eqref{gKeq}. 
Every point symmetry can be expressed in an equivalent, characteristic form 
\begin{equation}
\mathrm{\hat X} =P\partial_u, 
\quad
P=\eta(x,t,u)-\xi(x,t,u)u_x-\tau(x,t,u)u_t
\end{equation} 
which acts only on $u$. 
Invariance of the equation \eqref{gKeq} is given by the condition 
\begin{equation}\label{symm-deteqn}
0=D_tP - a(t) D_x^5 P - b(t)D_x^3 P - c(t)D_x(f(u)P)
\end{equation}
holding for all solutions $u(x,t)$ of equation \eqref{gKeq}. 
This condition \eqref{symm-deteqn} splits with respect to $x$-derivatives of $u$, 
which yields a overdetermined system of equations 
on $\eta(x,t,u)$, $\xi(x,t,u)$, $\tau(x,t,u)$, along with $a(t)$, $b(t)$, $c(t)$, $f(u)$,
subject to the classification conditions \eqref{classify}. 
It is straightforward to set up and solve this determining system by Maple. 

We obtain the following results. 
The point symmetries admitted by the generalized Kawahara equation \eqref{gKeq} 
in the general case \eqref{classify} are generated by 
\begin{flalign}
&\label{symm1}
\tau = 0,
\quad
\xi = 1,
\quad
\eta = 0 .
&
\end{flalign}
All special cases for which additional point symmetries are admitted 
consist of:
\begin{subequations}
\begin{flalign}
&\label{symm2}
\tau = \alpha t +\beta,
\quad
\xi = \tfrac{1}{5} \alpha x,
\quad
\eta = 0 
&\\
&
f(u) \text{ arbitrary},
\quad
b(t) = (\alpha t+\beta) ^{-\frac{2}{5}},
\quad
c(t) = \gamma(\alpha t+\beta) ^{-\frac{4}{5}};
&
\end{flalign}
\end{subequations}
\begin{subequations}
\begin{flalign}
&\label{symm3}
\tau = \alpha t+\beta,
\quad
\xi = \tfrac{1}{5}\alpha x -\delta \gamma f_0{\textstyle\int} c(t)\,dt,
\quad
\eta = -(\alpha\delta/f_3)(u+f_2) 
&
\\& 
f(u) = f_1(u+f_2) ^{f_3}+f_0,
\quad
b(t) = (\alpha t+\beta)^{-\frac{2}{5}},
\quad
c(t) = \gamma(\alpha t+\beta)^{\delta-\frac{4}{5}},
\quad
f_1, f_3\neq 0;
&
\end{flalign}
\end{subequations}
\begin{subequations}
\begin{flalign}
&\label{symm4}
\tau =0,
\quad
\xi = {\textstyle\int} c(t)\, dt,
\quad
\eta = -1/f_1
&
\\&
f(u) = f_1 u+f_0,
\quad
b(t) \text{ arbitrary},
\quad
c(t) \text{ arbitrary},
\quad
f_1\neq 0;
&
\end{flalign}
\end{subequations}
\begin{subequations}
\begin{flalign}
&\label{symm5}
\tau = d(t)/D(t)^5,
\quad
\xi = \big(\tfrac{1}{5}-\delta +\delta \beta/D(t)\big)x,
\quad 
\eta = (\alpha\beta\delta^2/(\gamma^3f_1)) x  -\big(\tfrac{1}{5}-\delta\beta/D(t)\big)(u + f_0/f_1)
&
\\&
\begin{aligned}
& f(u) = f_1 u+f_0,
\quad
b(t) = \gamma^{2} d(t)^{-\frac{2}{5}}D(t)^2,
\quad
c(t) = \gamma^{3} d(t)^{\delta-1}D(t)^3,
\quad
f_1\neq 0, 
\\&
D(t)= \alpha d(t)^{\delta} + \beta, 
\quad
d'(t) = D(t)^5, 
\quad
\gamma,\delta\neq 0;
\end{aligned}
&
\end{flalign}
\end{subequations}
\begin{subequations}
\begin{flalign}
&\label{symm6}
\tau = 0,
\quad
\xi = {\textstyle\int} c(t)\, dt,
\quad
\eta = -(1/f_1)(u+f_2) 
&
\\&
f(u) = f_1\ln(u+f_2) +f_0,
\quad
b(t) \text{ arbitrary},
\quad
c(t) \text{ arbitrary},
\quad
f_1\neq 0.
&
\end{flalign}
\end{subequations}

These symmetries can be interpreted by considering the forms of generators for 
space translations $\mathrm{X}_\text{space} =\partial_x$, 
time translations $\mathrm{X}_\text{time} =\partial_t$, 
shifts $\mathrm{X}_\text{shift} =-\partial_u$, 
scalings $\mathrm{X}_\text{scal.} =pt\partial_t +qx\partial_x +ru\partial_u$
where $p,q,r$ are the weights, 
time-dependent dilations $\mathrm{X}_\text{dil.} =q(t)\partial_t +p(t)x\partial_x +r(t)u\partial_u$, 
and Galilean boosts $\mathrm{X}_\text{Gal.} =\big({\textstyle\int}\nu(t)\,dt\big)\partial_x$
where $\nu(t)$ is the relative speed. 

In particular, 
symmetry \eqref{symm1} represents 
a space-translation; 
symmetry \eqref{symm2} represents 
a time-translation ($\beta$) combined with a scaling ($\alpha$); 
symmetry \eqref{symm3} represents 
a time-translation ($\beta$) combined with the composition of 
a scaling ($\alpha$), a shift ($\delta\alpha f_2/f_3$), and a Galilean boost ($-\delta \gamma f_0$) with relative speed $c(t)$; 
symmetry \eqref{symm4} represents 
a shift ($1/f_1$) combined with a Galilean boost with relative speed $c(t)$; 
symmetry \eqref{symm6} represents 
a shift ($f_2/f_1$) combined with a scaling and a Galilean boost with relative speed $c(t)$; 
symmetry \eqref{symm5} represents the composition of 
an $x$-dependent shift and a time-dependent dilation.

\section{Conservation laws}

Conservation laws are of basic importance because 
they provide physical, conserved quantities for all solutions $u(x,t)$,
and they can be used to check the accuracy of numerical solution methods. 
A general discussion of conservation laws and their applications to differential equations
can be found in Refs.\cite{Olv,BluCheAnc}. 

A local conservation law for the generalized Kawahara equation \eqref{gKeq} 
is a continuity equation 
\begin{equation}\label{conslaw}
D_t T+D_x X=0
\end{equation}
holding for all solutions $u(x,t)$ of equation \eqref{gKeq}, 
where the conserved density $T$ and the spatial flux $X$ are
functions of $t$, $x$, $u$, and $x$-derivatives of $u$
(with $t$-derivatives of $u$ being eliminated through the equation \eqref{gKeq}). 
If $T=D_x\Theta$ and $X=-D_t\Theta$ hold for all solutions $u(x,t)$,
where $\Theta$ is some function of $t$, $x$, $u$, and $x$-derivatives of $u$, 
then the continuity equation \eqref{conslaw} becomes an identity. 
Conservation laws of this form are called locally trivial,
and two conservation laws are considered to be locally equivalent if they differ
by a locally trivial conservation law. 
The global form of a non-trivial conservation law is given by 
\begin{equation}\label{globalconslaw}
\frac{d}{dt}\int_{\Omega} T\, dx = -X\Big|_{\partial\Omega}
\end{equation}
where $\Omega\subseteq\mathbb{R}$ is any fixed spatial domain. 

Every local conservation law can be expressed in an equivalent, characteristic form 
(analogous to the evolutionary form for symmetries) \cite{Olv}
which is given by a divergence identity
\begin{equation}\label{char-eqn}
D_t \tilde T+D_x \tilde X= (u_t -a(t) u_{xxxxx} - b(t)u_{xxx} -c(t)f(u) u_x)Q
\end{equation}
holding off of the set of solutions of the generalized Kawahara equation \eqref{gKeq},
where $\tilde T= T+D_x\Theta$ and $\tilde X= X-D_t\Theta$ 
are a conserved density and spatial flux that are locally equivalent to $T$ and $X$, 
and where 
\begin{equation}\label{QTrel}
Q = E_u(\tilde T) 
\end{equation} 
is a function of $t$, $x$, $u$, and $x$-derivatives of $u$. 
This function is a called a multiplier \cite{Olv,AncBlu97,BluCheAnc}. 
Here $E_u$ denotes the Euler operator with respect to $u$ \cite{Olv}. 

For evolution equations, 
there is a one-to-one correspondence between non-zero multipliers 
and non-trivial conservation laws up to local equivalence \cite{Olv,AncBlu02},
and the conservation laws of basic physical interest arise from 
multipliers of low order \cite{Anc16a}
\begin{equation}\label{low-order-Q}
Q(t,x,u,u_x,u_{xx},u_{xxx},u_{xxxx}) . 
\end{equation}
Such multipliers correspond to conserved densities of the form 
$T(t,x,u,u_x,u_{xx})$ modulo a trivial conserved density. 
A function \eqref{low-order-Q} will be a multiplier iff 
$E_u((u_t -a(t) u_{xxxxx} - b(t)u_{xxx} -c(t)f(u) u_x)Q)=0$ holds identically. 
This condition splits with respects to the $x$-derivatives of $u$ that do not appear in $Q$.
The resulting overdetermined system consists of 
the adjoint of the symmetry determining equation \eqref{symm-deteqn}
\begin{equation}\label{adjsymm-deteqn}
0=-D_tQ + a(t) D_x^5 P + b(t)D_x^3 P + c(t)f(u)D_x Q
\end{equation}
holding for all solutions $u(x,t)$ of equation \eqref{gKeq},
plus the Helmholtz equations \cite{AncBlu02,Anc16a}
\begin{align}\label{helmholtz-deteqn}
Q_u = E_u(Q),
\quad
Q_{u_x} = -E_{u_x}^{(1)}(Q),
\quad
Q_{u_{xx}} = E_{u_{xx}}^{(2)}(Q),
\quad
Q_{u_{xxx}} = -E_{u_{xxx}}^{(3)}(Q)
\end{align}
which are necessary and sufficient for $Q$ to have the variational form \eqref{QTrel}. 
Here $E_v^{(1)}$, and so on, denote the higher Euler operators with respect to a variable $v$ \cite{Olv,Anc16a}. 

It is straightforward using Maple to set up and solve this determining system \eqref{adjsymm-deteqn}--\eqref{helmholtz-deteqn} 
for $Q(t,x,u,u_x,u_{xx},u_{xxx},u_{xxxx})$ along with $a(t)$, $b(t)$, $c(t)$, $f(u)$,
subject to the classification conditions \eqref{classify}. 

For each solution $Q$, a corresponding conserved density $T$ and spatial flux $X$
can be derived (up to local equivalence) 
by integration of the divergence identity \eqref{char-eqn}
\cite{BluCheAnc,Anc16a}. 
We obtain the following results. 

The multipliers and conserved densities admitted by the generalized Kawahara equation \eqref{gKeq} 
in the general case \eqref{classify} are linear combinations of 
\begin{flalign}
&
Q=1,
\quad
T= u;
&
\label{conslaw1a}
\\
&
Q=u, 
\quad
T= \tfrac{1}{2}u^2 .
\label{conslaw1b}
&
\end{flalign}
All special cases for which additional multipliers and conserved densities are admitted 
consist of:
\begin{subequations}
\begin{flalign}
&
Q=u_{xxxx} +\alpha u_{xx} +\beta {\textstyle\int} f(u)\,du, 
\quad
T= \tfrac{1}{2}u_{xx}^2-\tfrac{1}{2}\alpha u_x^2 +\beta \big( u{\textstyle\int} f(u)\, du - {\textstyle\int} uf(u)\, du \big)
\label{conslaw2}
&\\
& 
f(u) \text{ arbitrary},
\quad
b(t) = \alpha, 
\quad
c(t) = \beta ;
&
\end{flalign}
\end{subequations}
\begin{subequations}
\begin{flalign}
&
Q=(\alpha t+\beta)u_{xxxx} +(\alpha t+\beta)^{\frac{3}{5}} u_{xx} 
+\tfrac{1}{5}\alpha x(u+f_2) +(\gamma/(f_3+1))(\alpha t+\beta)^{\frac{1}{5}(f_3+1)}(u+f_2)f(u),
&\\
&
T= \tfrac{1}{2}(\alpha t+\beta)u_{xx}^2-\tfrac{1}{2}(\alpha t+\beta)^{\frac{3}{5}}u_x^2 
+\tfrac{1}{10}\alpha x (u+f_2)^2
+(\gamma/(f_3+1)) (\alpha t+\beta)^{\frac{1}{5}(f_3+1)}\big( {\textstyle\int} (u+f_2)f(u)\,du \big)
\label{conslaw3}
&\\
& 
f(u) =f_1(u+f_2)^{f_3} +f_0, 
\quad
b(t) = (\alpha t+\beta) ^{-\frac{2}{5}},
\quad
c(t) = \gamma(\alpha t+\beta) ^{\frac{1}{5}(f_3 -4)},
\quad
f_4\neq 0;
\label{conds3}
&
\end{flalign}
\end{subequations}
\begin{subequations}
\begin{flalign}
&
Q=\big({\textstyle\int} c(t)\,dt\big) (f_1 u+f_0) +x,
\quad
T= \tfrac{1}{2} \big({\textstyle\int} c(t)\,dt\big) (f_1 u^2+2f_0 u)  +xu
\label{conslaw4}
&\\
& 
f(u) =f_1u +f_0, 
\quad
b(t) \text{ arbitrary},
\quad
c(t) \text{ arbitrary};
\label{conds4}
&
\end{flalign}
\end{subequations}
\begin{subequations}
\begin{flalign}
&
\begin{aligned}
& Q=
25\gamma f_1\big( d(t)u_{xxxx} + d(t)^{\frac{3}{5}}u_{xx} +\tfrac{1}{5} x D(t) u \big)
+\tfrac{1}{2}\alpha x^2 +\alpha f_0 x{\textstyle\int} c(t)\,dt  
\\&\qquad\quad
+\tfrac{25}{2}\gamma^2 f_1f_0 d(t)^{\frac{2}{5}} uf(u)
+\tfrac{1}{2}\alpha f_0 \big( {\textstyle\int} c(t)\,dt \big)^2 f(u)
\end{aligned}
&\\
&
\begin{aligned}
& T= \tfrac{25}{2}\gamma f_1\big( d(t)u_{xx}^2 - d(t)^{\frac{3}{5}}u_{x}^2 +\tfrac{1}{5} x D(t) u^2 \big)
+\tfrac{1}{2}\alpha x^2 u +\alpha f_0 x \big({\textstyle\int} c(t)\,dt\big) u
\\&\qquad\quad
+\tfrac{25}{2}\gamma^2 f_1f_0 d(t)^{\frac{2}{5}} {\textstyle\int} uf(u)\,du 
+\tfrac{1}{2}\alpha f_0 \big( {\textstyle\int} c(t)\,dt \big)^2 {\textstyle\int} f(u)\,du 
\end{aligned}
\label{conslaw5}
&\\
& 
\begin{aligned}
& f(u) =f_1u +f_0, 
\quad
b(t) = d(t)^{-\frac{2}{5}}, 
\quad
c(t) =\gamma d(t)^{-\frac{3}{5}}, 
\\&
d'(t) = D(t)^{\frac{1}{2}},
\quad 
D(t) = \alpha d(t)^{\frac{2}{5}}+\beta ;
\end{aligned}
\label{conds5}
&
\end{flalign}
\end{subequations}

The physical meaning of these conservation laws 
can be seen by considering their global form \eqref{globalconslaw}. 

For general $f(u)$, 
the three admitted conservation laws \eqref{conslaw1a}, \eqref{conslaw1b}, \eqref{conslaw2} 
respectively yield the conserved integrals 
\begin{equation}\label{globalconslaw1,2}
C_1 = \int_{\Omega} u\, dx,
\quad
C_2 = \int_{\Omega} \tfrac{1}{2}u^2\, dx,
\quad
C_3 = \int_{\Omega} \Big( 
\tfrac{1}{2}u_{xx}^2-\tfrac{1}{2}\alpha u_x^2 +\beta \big( u{\textstyle\int} f(u)\, du - {\textstyle\int} uf(u)\, du \big) 
\Big)\, dx
\end{equation}
These represent the mass, the $L^2$-norm, and the gradient-energy for solutions $u(x,t)$. 
When $f(u)$ has a power-law form \eqref{conds3}, 
the conservation law \eqref{conslaw3} yields the conserved integral 
\begin{equation}\label{globalconslaw3}
C_4 = \int_{\Omega} \Big( 
\tfrac{1}{2}(\alpha t+\beta)u_{xx}^2-\tfrac{1}{2}(\alpha t+\beta)^{\frac{3}{5}}u_x^2 
+\tfrac{1}{10}\alpha x (u+f_2)^2
+(\gamma/(f_3+1)) (\alpha t+\beta)^{\frac{1}{5}(f_3+1)}\big( {\textstyle\int} (u+f_2)f(u)\,du \big)
\Big)\,dx
\end{equation}
which represents a dilational Galilean energy for solutions $u(x,t)$. 
When $f(u)$ is a linear polynomial \eqref{conds4} and \eqref{conds5}, 
the two admitted conservation laws \eqref{conslaw4} and \eqref{conslaw5}
yield the respective conserved integrals 
\begin{equation}\label{globalconslaw4}
C_5 = \int_{\Omega} \Big( \tfrac{1}{2} C(t)\big( f_2 u^2+2f_1 u\big) +xu \Big) \,dx
\end{equation}
which represents a Galilean momentum, 
and
\begin{equation}\label{globalconslaw5}
\begin{aligned}
C_6 = & \int_{\Omega} \Big( 
\tfrac{25}{2}\gamma f_1\big( d(t)u_{xx}^2 - d(t)^{\frac{3}{5}}u_{x}^2 +\tfrac{1}{5} x D(t) u^2 \big)
+\tfrac{1}{2}\alpha x^2 u +\alpha f_0 x \big({\textstyle\int} c(t)\,dt\big) u
\\&\qquad\quad
+\tfrac{25}{2}\gamma^2 f_1f_0 d(t)^{\frac{2}{5}} {\textstyle\int} uf(u)\,du 
+\tfrac{1}{2}\alpha f_0 \big( {\textstyle\int} c(t)\,dt \big)^2 {\textstyle\int} f(u)\,du 
\Big)\,dx,
\end{aligned}
\end{equation}
which represents a generalized dilational Galilean energy-momentum, 
where $C(t)={\textstyle\int} c(t)\,dt$. 

These interpretations are reinforced by the Hamiltonian symmetries 
associated to the conserved integrals, 
as discussed in the next section.

\section{Connection between conservation laws and symmetries}

The generalized Kawahara equation \eqref{gKeq} has a Hamiltonian structure, 
on any fixed spatial domain $\Omega\subseteq\mathbb{R}$, 
which is given by 
\begin{equation}\label{hamilstruc}
u_t = {\mathcal H}(\delta H/\delta u), 
\quad
H = \int_{\Omega} ( \tfrac{1}{2} ( a(t) u_{xx}{}^2 + b(t) u_x{}^2 ) - c(t) F(u) )\,dx, 
\quad
F = {\textstyle\int\!\int} f\, du\, du = u{\textstyle\int} f\, du - {\textstyle\int} uf\, du 
\end{equation}
where the Hamiltonian operator \cite{Olv} is a total $x$-derivative 
\begin{equation}\label{Hop}
{\mathcal H} = D_x .
\end{equation}
This Hamiltonian structure yields a corresponding Lagrangian 
\begin{equation}\label{lagr}
L = \tfrac{1}{2} ( {-}v_t v_x + a(t) v_{xxx}{}^2 + b(t) v_{xx}{}^2 ) - c(t) F(v_x), 
\quad
u=v_x
\end{equation}
when a potential is introduced. 
In potential form, the generalized Kawahara equation \eqref{gKeq} is an Euler-Lagrange equation
\begin{equation}\label{gKeq-pot} 
E_v(L) = v_{tx} - a(t) v_{xxxxxx} - b(t)v_{xxxx} - c(t)f(v_x) v_{xx} 
= u_t - a(t) u_{xxxxx} - b(t)u_{xxx} - c(t)f(u) u_x . 
\end{equation}

For any Hamiltonian evolution equation, 
there is a correspondence \cite{Olv} 
that produces a symmetry from each admitted conservation law. 
This correspondence is a Hamiltonian analog of Noether's theorem. 
It can be formulated for the generalized Kawahara equation \eqref{gKeq} 
by the explicit relation 
\begin{equation}\label{PQmap}
P = {\mathcal H}(\delta C/\delta u) = D_x Q
\end{equation} 
involving the characteristic function $P$ of the symmetry generator $\mathrm{\hat X} =P\partial_u$ 
and the multiplier $Q$ associated to the conserved integral 
$C = {\textstyle \int}_{\Omega} T\, dx$
given by a local conservation law \eqref{conslaw}. 
This correspondence is one way: every conservation law yields a symmetry. 
The converse holds iff the symmetry has the Hamiltonian form \eqref{PQmap}, 
which requires that $E_u(P)=0$. 

The same correspondence \eqref{PQmap} can be derived from Noether's theorem 
applied to the Lagrangian for the generalized Kawahara equation in potential form \eqref{gKeq-pot} as follows. 
Every local conservation law \eqref{conslaw} 
admitted by the generalized Kawahara equation 
arises from a multiplier $Q$. 
When the multiplier is expressed in terms of the potential, $Q\big|_{u=v_x}=Q^v$, 
it yields a local conservation law for the equation in potential form. 
From Noether's theorem, 
there is one-to-one correspondence between 
the local conservation laws (up to equivalence)
and the variational symmetries of this equation \eqref{gKeq-pot}. 
This correspondence is explicitly given by $P^v = Q^v$,
where $P^v$ is the characteristic function of the variational symmetry 
$\mathrm{\hat Y} = P^v\partial_v$
and $Q^v$ is multiplier of the conservation law. 
Through the relation $u=v_x$, 
the prolongation of the variational symmetry 
$\mathrm{pr}\mathrm{\hat Y} = \mathrm{\hat Y} + D_x P^v\partial_{v_x}$
yields a symmetry of the generalized Kawahara equation itself \eqref{gKeq},
as given by $P\big|_{u=v_x}=D_x P^v$. 
This symmetry relation combined with $P^v = Q^v = Q\big|_{u=v_x}=Q^v$ 
is precisely the correspondence \eqref{PQmap}. 

The set of Hamiltonian symmetries 
\begin{equation}
\mathrm{\hat X}_\text{Ham.}=D_x Q\partial_u,
\end{equation}
or equivalently the set of variational symmetries 
\begin{equation}
\mathrm{\hat Y}_\text{var.} = Q\big|_{u=v_x}\partial_v,
\end{equation}
will be a Lie subalgebra of the Lie algebra of symmetries 
admitted by the generalized Kawahara equation \eqref{gKeq}. 
From the conserved integrals \eqref{globalconslaw1,2}--\eqref{globalconslaw5}, 
we find that 
$C_2$ produces a space-translation, 
$C_3$ produces a time-translation, 
$C_4$ produces a scaling combined with a Galilean boost, 
$C_5$ produces a shift combined with a Galilean boost, 
and $C_6$ produces a Galilean boost combined with 
an $x$-dependent shift and a time-dependent dilation.

\section{Concluding remarks}

We have presented a complete classification of 
all low-order conservation laws and all point symmetries 
admitted by the generalized Kawahara equation \eqref{gKeq}. 
This classification includes as a special case the usual Kawahara equation \eqref{Keq}. 

The symmetries can be used to obtain exact group-invariant solutions,
while the conservation laws can be used to investigate these solutions 
as well as to study the initial-value problem. 

We have also explained a general connection between conservation laws and symmetries
for the generalized Kawahara equation
through the Hamiltonian structure of this equation 
and its relationship to Noether's theorem using a potential formulation.

\nocite{*}
\bibliographystyle{aipnum-cp}%

\begin{thebibliography}{0}

 \bibitem{AncBlu97}
S.C. Anco and G. Bluman,
Direct Construction of Conservation Laws from Field Equations,
{\em Phys. Rev. Lett.} 78 (1997), 2869--2873.

\bibitem{AncBlu02}
S. C. Anco and G. Bluman,
Direct construction method for conservation laws of partial differential equations Part II: General treatment,
{\em Euro. J. Appl. Math.} 41 (2002), 567--585.

\bibitem{Anc16a}
S.C. Anco,
Generalization of Noether's theorem in modern form to non-variational partial differential equations.
To appear in {\em Fields Institute Communications: Recent progress and Modern Challenges in Applied Mathematics, Modeling and Computational Science};
arXiv: mathph/1605.08734 (2016).

\bibitem{AncBluWol}
S.C. Anco, G. Bluman, and T. Wolf, 
Invertible mappings of nonlinear PDEs to linear PDEs through admitted conservation laws, 
{\em Acta Appl. Math.} 101 (2008), 21--38. 

\bibitem{BluCheAnc}
G.W. Bluman, A Cheviakov, S.C. Anco,
Applications of Symmetry Methods to Partial Differential Equations.
\emph{New York: Springer} (2009).

\bibitem{FreSam}
I.L. Freire, J.C.S. Sampaio,
Nonlinear self-adjointness of a generalized fifth-order KdV equation,
{\em J. Phys. A: Math. Theor.} 45 (2012), 032001 (7pp). 

\bibitem{Has}
H. Hasimoto, 
Water waves, 
{\em Kagaku}, 40 (1970), 401--408.

\bibitem{KakOno}
T. Kakutani and H. Ono, 
Weak non-linear hydromagnetic waves in a cold collision free plasma, 
{\em J. Phys. Soc. Japan} 26 (1969), 1305--1318.

\bibitem{Kak}
T. Kakutani, 
Effect of an uneven bottom on gravity waves, 
{\em J. Phys. Soc. Japan} 30 (1971), 272--276.

\bibitem{Kaw}
T. Kawahara, 
Oscillatory solitary waves in dispersive media, 
{\em J. Phys. Soc. Japan}, 33 (1972), 260--264. 

\bibitem{LiuLiLiu}
H. Liu H, J. Li, L Liu, 
Lie symmetry analysis, optimal systems and exact solutions to the fifth-order KdV types of equations,
{\em J. Math. Anal. Appl.} 368 (2010) 551--558

\bibitem{Olv}
P.J. Olver,
Applications of Lie Groups to Differential Equations.
\emph{Berlin: Springer}  (1986).

\bibitem{Zab}
N.J. Zabusky, 
A synergetic approach to problems of nonlinear dispersive wave propagation and interaction, 
in {\it Proc. Symp. Nonlinear Partial Differential Equations} (ed. W. Ames), 
Academic Press (1967), 223-258. 


\end{thebibliography}

\end{document}